\documentclass[apj]{emulateapj}
\newcommand{\tint}{{\rm T}_{\rm int}}
\newcommand{\teff}{{\rm T}_{\rm eff}}
\newcommand{\enstatite}{MgSiO$_3$}
\newcommand{\forsterite}{Mg$_2$SiO$_4$}

\shortauthors{Barman et al.}
\journalinfo{The Astrophysical Journal, {\em in press}}
\submitted{Received 2005 May 22; accepted 2005 June 27} 

\begin{document}

\title{Phase-Dependent Properties of Extrasolar Planet Atmospheres}
 \author{Travis S. Barman}
 \affil{Department of Physics and Astronomy, University of California at 
        Los Angeles, Los Angeles, CA 90095\\
        Email: {\tt barman@astro.ucla.edu}}
 \author{Peter H. Hauschildt}
 \affil{Hamburger Sternwarte, Gojenbergsweg 112, 21029 Hamburg, Germany\\
        Email: {\tt yeti@hs.uni-hamburg.de}}
 \author{France Allard}
 \affil{C.R.A.L (UML 5574) Ecole Normale Superieure, 69364 Lyon Cedex 7, France\\
        EMail: {\tt fallard@ens-lyon.fr}}

\begin{abstract} 
Recently the Spitzer Space Telescope observed the transiting extrasolar
planets, TrES-1 and HD209458b. These observations have provided the first
estimates of the day side thermal flux from two extrasolar planets orbiting
Sun-like stars.  In this paper, synthetic spectra from atmospheric models are
compared to these observations.  The day-night temperature difference is
explored and phase-dependent flux densities are predicted for both planets.
For HD209458b and TrES-1, models with significant day-to-night energy
redistribution are required to reproduce the observations.  However, the
observational error bars are large and a range of models remains viable.
\end{abstract}


\section{Introduction}

Of the more than 100 extrasolar planetary systems discovered so far, only 7
have near edge-on orbits.  These transiting planets are crucial for
understanding giant planets in general since their masses and radii can easily
be determined, and careful multi-wavelength observations can reveal some
information about the planet's atmosphere \cite[]{Brown01,
Charbonneau02,Vidal-Madjar03}.  Recently, two transiting planets, HD209458b and
TrES-1, were observed with the Spitzer Space Telescope, providing the first
direct measurements of their thermal flux.  By comparing IR fluxes in and out
of secondary eclipse (when the planet is behind the star), \cite{Tres1Spitzer}
measured the planet-star flux density ratio at 4.5 and 8 \micron\ for TrES-1.
Independently, \cite{HDSpitzer} measured the flux density ratio at 24 \micron\ for
HD209458b.  These measurements provide the best constraints, so far, on the
thermal structure and chemical composition of highly irradiated EGPs.  

HD209458b and TrES-1 both have very short periods (just a few days) and orbital
separations less than 0.05AU \cite[]{Henry00,TrES-1}.  At such small orbital
separations, they are substantially heated by radiation from their parent
stars.  Following the discovery of the planet around 51 Pegasi
\cite[]{Mayor95}, a variety of atmosphere models suitable for estimating the
properties of these so-called ``hot Jupiters" were published
\cite[]{Seager1998, Barman01, Sudarsky03}.  However, since these planets
probably have strong day-to-night photospheric differences, their potential
lack of symmetry adds complications to an already difficult model atmosphere
problem.  Several of the most challenging issues are the coverage and types of
clouds (if present), redistribution of the absorbed stellar flux by atmospheric
currents, depth dependent non-solar abundances, and photospheric temperature
and pressure gradients from the day to night sides.  These problems have been
dealt with (or avoided) in a variety of ways and, therefore, a variety of model
predictions exist.

New models, that estimate the horizontal atmospheric gradients under the
assumption of radiative-convective equilibrium, are presented below.  Several
of the standard assumptions for global energy redistribution are explored.
Model results are also compared to the most recent Spitzer measurements and
estimates are given for the planet-star flux density ratios in the Spitzer
24 \micron\ MIPS band and the four IRAC bands as a function of orbital phase.
 
\section{Model Construction}

The irradiated models presented below were calculated using the {\tt PHOENIX}
atmosphere code \cite[]{yeti99,Allard01} adapted to include extrinsic radiation
as described in Barman et al. (2001; here after BHA01, 2002)
\nocite{Barman01,Barman02}.  The spherically symmetric radiative transfer and
chemical equilibrium equations were solved self-consistently, while explicitly
accounting for the wavelength-dependent extrinsic radiation.  The extrinsic
radiation was also modeled with {\tt PHOENIX} and, in each case, closely matches
the observed parent star spectrum.

The major differences between the computation of the BHA01 models and those
presented here lie with the treatment of dust in the atmosphere and the
assumptions concerning the redistribution of absorbed stellar flux over the
planet's day and night hemispheres.

\begin{deluxetable}{lcll}
\tablecolumns{4}
\tablewidth{22pc}
\tablecaption{Model Parameters}
\tablehead{\colhead{}& \colhead{} & \colhead{HD209458b$^\dagger$}&\colhead{TrES-1$^\ddagger$}}
\startdata
T$_{\star}$ (K)         &\nodata & $6088  \pm 56$         & $5250 \pm 75$           \\
R$_{\star}$ (R$_\odot$) &\nodata & $1.145 \pm 0.049$      & $0.83 \pm 0.05$         \\
M$_{\star}$ (M$_\odot$) &\nodata & $1.06$                 & $0.89$                  \\
                        &        &                        &                         \\
R$_{p}$ (R$_{\rm Jup}$) &\nodata & $1.42^{+0.10}_{-0.13}$ & $1.04^{+0.08}_{-0.05}$  \\
M$_{p}$ (M$_{\rm Jup}$) &\nodata & $0.69$                 & $0.76$                  \\
a (AU)                  &\nodata & 0.0468                 & 0.0393                  \\
\enddata
\label{tab1}
\tablerefs{
  ($\dagger$) \cite{Ribas03}, \cite{Cody02}; 
  ($\ddagger$) \cite{Sozzetti04}}
\end{deluxetable} 

\subsection{Cloud-free Assumption}

Unlike many brown dwarfs, EGPs do not necessarily have convective photospheres.
For EGPs with small orbital separations, irradiation can suppress convection to
depths well below the photosphere, leading to a fully radiative atmosphere
across most of the day side \cite[]{Guillot96}.  Since radiative photospheres have
short sedimentation time-scales, cloud growth should be difficult to sustain
making the cloud-free assumption reasonable. However, this assumption may break
down if strong zonal winds are present and advective time-scales are comparable
to sedimentation time-scales.  The efficiency of gravitational settling will also
depend on the poorly constrained eddy diffusion coefficient \cite[]{Rossow78}.

In BHA01, cloud-free atmospheres were modeled using the ``Cond'' opacity setup
\cite[]{Allard01}.  The Cond setup accounts for dust formation in the
atmosphere, as determined by chemical equilibrium equations, but excludes the
dust opacity when computing the thermal and spectroscopic properties of the
atmosphere.  Excluding the dust opacity was intended to approximate the effects
of cloud formation followed by efficient gravitational settling (often referred
to as ``Rainout'') which acts to deplete an atmosphere of many important
refractory elements.  Rainout and the depletion of refractory elements have
been recognized for some time as important processes in the atmospheres of
Jovian planets and brown dwarfs \cite[]{Fegley1996, Lodders1999, Burrows2000,
Marley2002}.  The success of these models is strong motivation for exploring
the rainout assumption in extrasolar planet atmospheres.  
 
While the Cond approximation does effectively remove refractory elements, it
does not alter the overall abundance of an element sequestered by grain
formation at a given temperature and pressure.  Grain formation and efficient
gravitational settling alter the abundance of an element by continually
removing the refractory elements that make up a grain until the grain is no
longer able to form due to a lack of one or more constituents. In the absence
of replenishment (e.g., by convective updrafts), certain layers of an
atmosphere where only gravitational settling has occurred will not only be free
of dust species but also free of a significant number of metals and related
molecules.  For a description of this effect in Jupiter's atmosphere, see
\cite{Lodders1999}.

The Cond case, therefore, does not go far enough in removing refractory
elements from the upper atmosphere and can lead to a small concentration of key
absorbers like TiO and VO that are important in irradiated atmospheres.  These
two molecules have strong absorption bands near the peak flux densities of
solar type stars.  Consequently, their presence or absence can greatly affect
the depth at which the stellar flux is absorbed in the planet's atmosphere and,
thus, alter the predicted atmospheric structure \cite[]{Hubeny03}.  

In this work, an improved cloud-free model was used that iteratively reduces
(at a given layer) the elemental abundances involved in grain formation and
recomputes the chemical equilibrium with each new set of stratified elemental
abundances.  This model is similar to the Rainout model of \cite{Burrows1999},
except that the depletion of elements is continued until grains (and thus grain
opacities) are no longer present (see also \cite{Allard03} for more details).
The resulting equilibrium chemistry and opacity sampling of this cloud-free
model are fully self-consistent, unlike the earlier Cond models which simply
excluded the grain opacity.  Also, for the models described below, Ti and V
were significantly depleted from the photosphere by the rainout process leading
to negligible concentrations of TiO and VO.  For a detailed discussion of the
differences between Rainout and Cond irradiated models, see Barman et al.
(2005, in preparation). 

\begin{figure}
\plotone{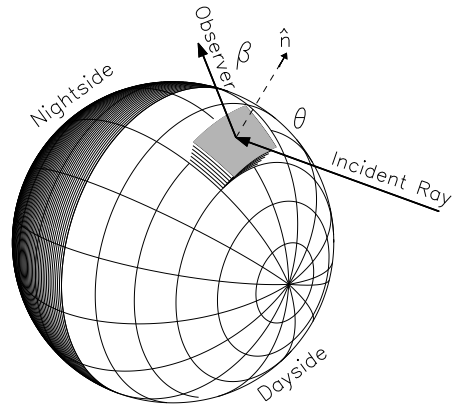}
\caption{Illustration of the planet's hemisphere divided into small regions,
each to be modeled independently.  Each concentric band around the sub-stellar
point (depicted here as the convergence point of longitudinal lines) receives
the same amount of incident stellar flux at the same incident angle.
Therefore, each band is assumed to have the same atmospheric structure and
emergent intensities.  The angle between the incident ray and the outward
pointing surface normal is $\theta$.  When computing the emergent flux, only
those intensities aligned with the observer line-of-sight are integrated over
the visible hemisphere.  The angle between the surface normal and an observer's
line-of-sight is $\beta$.
\label{toon1}}
\end{figure}

\subsection{Modeling the Day-Night Gradients}

The majority of static atmosphere models arrive at a single
temperature-pressure (T-P) profile intended to represent an average of either
the day side or over the entire planet \cite[]{Seager1998, Gouken2000,
Barman01, Sudarsky03}.  However, if most short period planets are well
represented by a static atmosphere in radiative-convective equilibrium, then
one should expect them to have a horizontal (day-to-night) temperature gradient
-- simply due to the center-to-limb variation in the amount of incident stellar
flux received by the planet.  In the absence of a 3-D model atmosphere code,
one approach that comes closer to the real solution is to divide a planet's day
side into a series of concentric regions around the sub-stellar point.  In the
static case, suitably large regions should interact very little via radiative
transfer processes, except perhaps near the terminator.

The planetary atmospheres described below were modeled by dividing the day side
into 10 concentric regions defined by $\mu = \cos(\theta)$, where $\theta$ is
the angle between the surface normal and the direction to the star (see Fig.
\ref{toon1}).  For these regions, $\mu$ ranged from 1.0 (at the sub-stellar
point) to 0.1 (the model region closest to the terminator), in steps of
$\Delta\mu = 0.1$.  The corresponding T-P profiles and emergent intensities
were modeled using 1-D, spherically symmetric, atmospheres each receiving
incident stellar flux along the appropriate angle for a given region.  For
these day side models, the radiative transfer equation was modified so that the incident
specific intensities along any $\mu$ and azimuthal angle $\phi$ were given by,
\begin{equation}
I_{inc,\lambda}(\mu,\phi) = I_{\circ,\lambda} \delta (\mu - \mu^\prime) \delta (\phi - \phi^\prime),
\end{equation}
with $\delta$ being the Dirac delta function.  In which case, it follows that
the incident fluxes are simply,
\begin{equation}
F_{inc,\lambda} (\mu) = \mu I_{\circ,\lambda}  = 
  \mu \left(\frac{R_{\star}}{d}\right)^2  F_{\star,\lambda} ,
\label{finc}
\end{equation}
where $F_{\star,\lambda}$ are the monochromatic fluxes from the star's surface,
$R_{\star}$ is the stellar radius, and $d$ is the distance from the stellar
surface to the planet's atmosphere.  For the night side, a single,
non-irradiated, model was used.  

All models were solved self-consistently so that each $\mu$-region had a
chemistry characterized by its T-P profile. By having chemical equilibria
consistent with the T-P profiles across the planet's atmosphere, this approach
naturally leads to variations in the important photospheric opacity sources
from the day to night side -- an important aspect when computing the synthetic
spectra.

Since EGPs are believed to have fully convective interiors, the {\em intrinsic}
effective temperature ($\tint$)
\footnote{
In the present work, $\tint$ characterizes the intrinsic luminosity of an
irradiated model atmosphere, defined by $4\pi R_{p}^{2} \sigma \tint^4$.  For
non-irradiated models, the normal $\teff$ is used to described the
flux and luminosity.}
for each modeled region was adjusted so that, after convergence, all T-P
profiles reached the same adiabat below the photosphere\footnote{The region
referred to as the photosphere lies roughly between $\rm P = 0.01$ and 1 bar,
corresponding to where the optical depth at IR wavelengths is near unity.}.  The
adiabat was selected based on planetary interior and evolution calculations for
a given mass, age, metallicity, and irradiation \cite[]{Baraffe03, Baraffe04,
Chabrier04}.  This entropy matching technique has also been used for irradiated
binary stars and allows one to assign models with different intrinsic
luminosities to different regions of the same star or planet
\cite[]{VazNordlund1985, NordlundVaz1990, Barman04}.  
 
The monochromatic fluxes from the model planet were obtained by integrating the
emergent intensities along an observer's line-of-sight ($los$) for a given
observer-planet-star orientation.  
\begin{equation}
F_{los,\lambda} = \int\limits_{los} {I_\lambda}(\theta,\phi) d\Omega  .
\label{fint}
\end{equation}
The integration was performed by distributing $\sim 2000$ points over the
entire surface with corresponding cubature integration weights ($w_i$).  The
distribution of points was determined by the minimization of potential energy
on the unit sphere.  This distribution is nearly orientation independent,
unlike the standard latitude-longitude grid, and results in very small
integration errors \cite[]{Steinacker96,Sloan01}.  Each point on the
observer-facing hemisphere was assigned an emergent intensity spectrum
corresponding to the angle, $\beta$, between the observer's $los$ and the
surface normal (see Fig. \ref{toon1}), and depending on the day or night region
in which the point belonged.  Numerical integration of Eq. \ref{fint} becomes a
sum over the visible points,
\begin{equation}
F_{los,\lambda} = \sum\limits_{i (visible)} w_i \mu_i I_\lambda(\mu_i) .
\label{fsum}
\end{equation}
In Eq. \ref{fsum}, $\mu = cos(\beta)$ and is not to be confused with the 10
$\mu$ values used to divide the day hemisphere.  With this technique, the
phase-dependent spectra can be estimated while taking into account the
center-to-limb variation of the planet's emergent intensities -- which might
include a combination of limb brightening and darkening.  The emergent
intensities were sampled over 114 directions per $\mu$-region and were assumed
to be azimuthally symmetric about the surface normal.  Assuming azimuthal
symmetry is justified since the present work is concerned primarily with the
thermal flux, not scattered light.  Phase-dependent optical spectra will be
explored in a later paper.

The simple approach outlined above has several limitations.  Since radiation
passing through one region into another is neglected, heating of the upper
atmosphere near the terminator may be underestimated at low gas pressures.
However, this is unlikely to affect the emergent thermal IR spectrum that forms
deeper in the atmosphere.  It has also been assumed that the planet is tidally
locked and, hence, presents a constant face to the parent star.  The models are
also time-independent and static and, consequently, neglect the effects of
zonal winds that could change the thermal profiles by coupling the hot day side
to the cooler night side.

\begin{figure}
\plotone{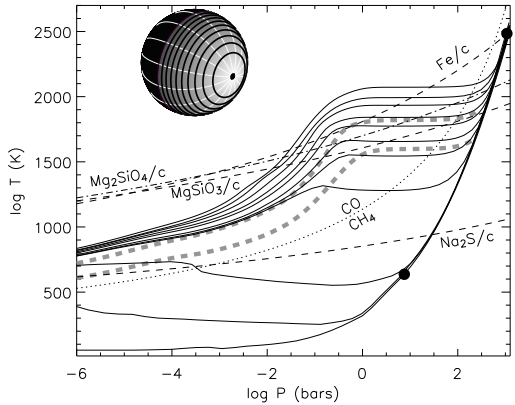}
\caption{Temperature versus pressure for a sequence of irradiated atmospheres.
For each model, the parameters for TrES-1 from Table \ref{tab1} were chosen and
only the direction of the incident flux relative to the surface normal was
varied.  From top to bottom, the models have $\mu = 1.0$ to $\mu = 0.1$ in
steps of 0.1.  The dashed lines indicate the approximate condensation curves
for three common grain species.  The dotted line indicates where gaseous CO and
CH$_4$ concentrations are equal (CO is dominant to the left of this line).  The
approximate regions represented by the collection of T-P profiles are shown as
solid black lines on the illustrative sphere.  The top-most T-P profile
corresponds to the sub-stellar point (black dot on the sphere).  The terminator
and night side (black hemisphere) are modeled with the non-irradiated profile
(lowest T-P curve).  The radiative-convective boundary at the sub-stellar point
and on the night side are labeled with filled circles.  The thick, grey, dashed
lines are T-P profiles for $\alpha = 0.5$ (top) and $\alpha=0.25$ (bottom)
models.
\label{tr1}}
\end{figure}

\subsection{Energy Redistribution}

The approach outlined in the previous section is designed to model the planet's
atmosphere under the assumption that the gas is truly static and in
radiative-convective equilibrium.  As such, this approach will predict the
maximum heating at the sub-stellar point and very little heating at the
terminator.  However, as mentioned above, an important consequence of stellar
heating is horizontal atmospheric flows capable of transporting appreciable
amounts of energy to the night side.  The impacts of horizontal motion on EGP
atmospheres have been modeled by a variety of groups, each predicting some
level of atmospheric circulation that depends strongly on adopted opacities and
general approaches to the problem \cite[]{Showman02,Cho03,Burkert05,Cooper05}.
Despite the differences in methods and results, the general consensus from
these hydrodynamic simulations is that circulations can redistribute a fraction
of the incident energy over large portions of a strongly irradiated planetary
atmosphere.

For single, 1-D, model atmospheres designed to reproduce the detailed
chemistry, opacities, and emergent spectrum (but not the atmospheric motions),
the effects of energy redistribution have been folded into a single parameter,
referred to as $\alpha$ in this work.  The $\alpha$ parameter is simply the
ratio of the planet's cross-sectional area ($\pi R_{p}^{2}$) and the surface
area of the planet from which the absorbed stellar luminosity is to be
re-emitted.  In this case, the stellar flux incident at the top of the model
atmosphere becomes,
\begin{equation}
 F_{\rm inc,\lambda} =
\alpha \left(\frac{R_{\star}}{d}\right)^2  F_{\star,\lambda}.
\label{eq1}
\end{equation}
For a more detailed description of the energy balance in an irradiated binary
companion and the development of a similar $\alpha$ parameter, see
\cite{paczynski80}.  
 
\begin{figure}
\plotone{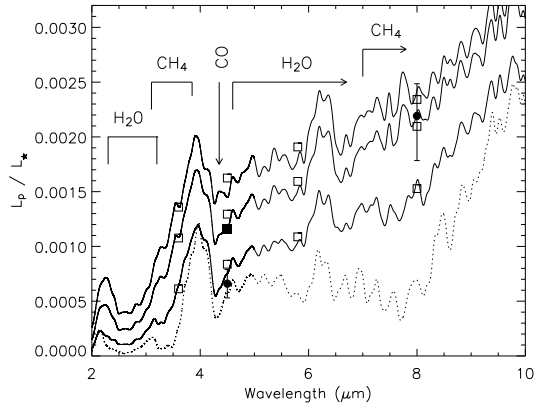}
\caption{ Planet-star flux density ratios for day side models assuming no
redistribution (top curve) and redistribution models with $\alpha = 0.5$
(middle solid curve) and $\alpha = 0.25$ (bottom solid curve).  The model T-P
profiles are shown in Fig. \ref{tr1}.  IRAC band fluxes for each model (found
by convolving with the IRAC response curves) are indicated with open squares
and filled circles show the Spitzer data with 1$\sigma$ error bars.  The
4.5 \micron\ IRAC value for a 10$\times$ solar, $\alpha = 0.5$, model is also
shown (solid square).  The lower dotted line corresponds to an isolated brown
dwarf model with $\teff = 1150$ K.
\label{tr2}}
\end{figure}

When the incident flux is scaled by $\alpha$, the underlying assumption is that
dynamical processes in the atmosphere are efficient enough to uniformly
distribute the incident luminosity over either the day hemisphere ($\alpha =
0.5$) or the entire sphere ($\alpha = 0.25$).  In addition, every point on
the day side (or entire surface) is assumed to be identical and, thus, can be
described by a single 1-D model with the same emergent and incident flux.  Note
that the $\alpha = 0.5$ case corresponds to the average (over $\mu$) of the
incident flux defined in Eq. \ref{finc}.  Also, the no-redistribution and $\alpha =
0.5$ cases receive and re-radiate the same amount of incident luminosity from
the day side, but will predict very different T-P profiles and phase-dependent
spectra (see below).

\section{Results}

The greatest number of observational constraints exists for the planets TrES-1
and HD209458b.  Models specifically tailored for these two objects are presented
below, adopting the parameters listed in Table \ref{tab1}, and are compared to
the recent Spitzer data.

\subsection{TrES-1}

Figure \ref{tr1} shows the sequence of solar metallicity T-P profiles across
the day and night sides of TrES-1.  On the day side, $\tint \sim 100$K which is
motivated by evolution calculations that reproduce the observed radius of
TrES-1 \cite[]{Baraffe05}.  The night side atmosphere model with the same
adiabat as the day side, has $\teff = 225$K based on the entropy matching
criteria mentioned above.  The sphere in Fig. \ref{tr1} shows the concentric
regions around the sub-stellar point represented by each atmosphere model.

As expected, a steep temperature gradient along an isobar, $\nabla T_{\rm P} =
(\partial T/\partial \mu)_{\rm P}$, is present from the sub-stellar point to
the terminator.  At $\rm P > 0.01$ bar, $\nabla T_{\rm P}$ is very large across
most of the day side and increases dramatically as the temperatures drop off
near the terminator (designated by the night side T-P profile).  Unlike the
deeper layers, the top most layers of the atmosphere receive a steady supply of
incident stellar flux, even when $\alpha > 60^\circ$ -- however at shallow
angles with respect to the surface normal.  Consequently, $\nabla T_{\rm P}$ is
smaller for $\rm P < 0.01$ bar compared to higher pressure depths for most of the
day side except near the terminator.  Note also that the T-P profiles (for
large $\alpha$) become flatter at low P and steeper at high P and, eventually,
become inverted very near the terminator.  

\begin{figure}
\plotone{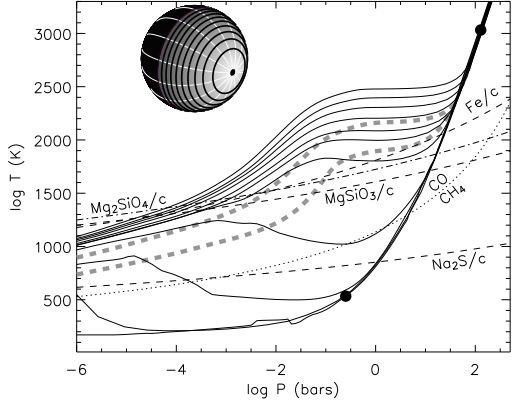}
\caption{
Same as Fig. \ref{tr1} but adopting the parameters for HD209458b in Table
\ref{tab1}.
\label{hd1}}
\end{figure}

The stellar heating also causes the atmospheric extension \footnote{The
atmospheric extension is defined here as the difference between the radius at
$\rm P = 10^{-6}$ bar and at $\rm P = 100$ bar.} to increase from the
terminator to the sub-stellar point by a factor of $\sim 5$.  The change in
extension coincides with a significant pressure gradient below the photosphere
along constant-height surfaces where the change in pressure from the day to
night side can be factors as large as 50 to 100.  Above the photosphere, the
day-to-night pressure gradient is present, but fairly small.  Despite the large
increase in extension, the change in radius is modest leading to no more than a
5\% increase in the area of an isobaric surface on the day side compared to the
night.  Also of interest is the radiative-convective boundary which is not on
an isobar and is significantly deeper at the sub-stellar point ($\rm P \sim
10^3$ bar) compared to the terminator and night side ($\rm P \sim 10$ bar).  

The condensation curves for Fe, \forsterite, \enstatite\ and Na$_2$S are also
shown in Fig. \ref{tr1}.  Cloud formation is typically believed to occur near
the intersection of the T-P profile and the condensation curve for a given
species.  While the models presented here are cloud-free, the condensation
curves suggest that photospheric clouds might be possible at a variety of
atmospheric depths and compositions but, in the no-redistribution case, would
be confined to $\sim$ 50\% of the day side around the sub-stellar point.  The
no-redistribution model also suggests that clouds might form at much greater
heights (i.e. lower P) around the sub-stellar point compared to the $\alpha =
0.5$ redistribution model.  The presence of clouds would have an impact on the
predicted T-P profile and, therefore, needs to be treated self-consistently.
Cloud formation across the day and night sides will be explored in a future
paper.

The high day side temperatures lead to an atmospheric chemistry dominated by
H$_2$, He, H$_2$O and CO.  However, near the terminator and on the night side,
most of the carbon is bound in CH$_4$.  In Fig. \ref{tr2}, the planet-star flux
density ratios\footnote{The ratio $(R_p/R_\star)^2$, determined from the
transit light curves, has error bars much smaller than the individual planet
or stellar radii.  Therefore, the flux density ratio comparison is not limited
by the typical uncertainties of an absolute flux determination.},
\begin{equation}
 \frac{L_p}{L_\star} = \left(\frac{R_p}{R_\star}\right)^2 \frac{F_p}{F_\star},
\end{equation}
for the no-redistribution model are compared to the Sptizer observations along
with standard 1-D models with $\alpha = 0.5$ and $\alpha = 0.25$.  Since the
stellar spectrum at IR wavelengths is fairly smooth, all of the features seen
are due to absorption in the planet's day side photosphere.  Strong absorption
by H$_2$O and CO are easily identified in the model.  The two hottest models (no
redistribution and $\alpha = 0.5$) are in good agreement with the observations
at 8 \micron\ but significantly overestimate the flux density at 4.5 \micron.
The models with complete redistribution ($\alpha = 0.25$) agrees reasonably well at
4.5 \micron\ and, at the 2$\sigma$ level, agrees with the 8 \micron\
observations.  Despite the broad wavelength span of the IRAC instrument, all of
the IRAC bands probe a fairly narrow region of the planetary atmosphere
between $\rm P = 10^{-1}$ and $10^{-2}$ bar.

\begin{figure}
\plotone{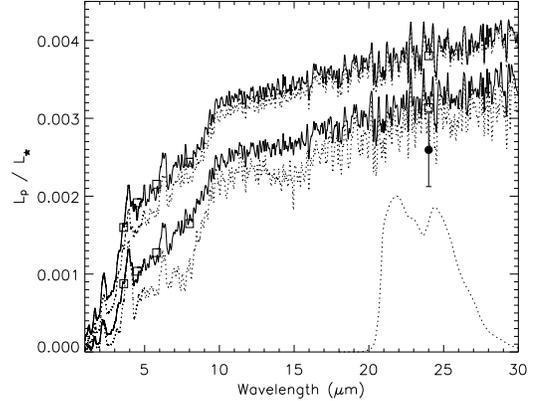}
\caption{Planet-star flux density ratio for HD209458b assuming no
redistribution (top solid curve) and redistribution models with $\alpha = 0.5$
(top dotted curve) and $\alpha = 0.25$ (bottom solid curve). The T-P profiles are
shown in Fig. \ref{hd1}.  Open squares indicate the predicted IRAC and MIPS
flux densities.  The observed 24 \micron\ MIPS value is shown with 1$\sigma$
error bars (filled circle) along with the 24 \micron\ MIPS response curve.
The lower dotted line is the flux density ratio for a Brown Dwarf with
$\teff = 1450$K.
\label{hd2}}
\end{figure}

Also shown in Fig. \ref{tr2} are the flux density ratios using a spectrum from
a non-irradiated brown dwarf model atmosphere with $\teff = 1150$K which
roughly corresponds to the equilibrium $\teff$ of the $\alpha = 0.25$ model.
While models of brown dwarfs appear to agree very well with recent Spitzer IRS
spectra of L and T dwarfs \cite[]{Roellig04}, the observations of TrES-1 are
clearly inconsistent with a standard brown dwarf spectrum, especially around 8
\micron.  This disagreement confirms that irradiated EGPs have atmospheric
structures very different from isolated brown dwarfs\footnote{\cite{Fortney05}
have also commented on the less than satisfactory agreement between models and
the EGP Spitzer data in light of the very good agreement between models and
Spitzer IRS brown dwarf spectra.}; a property that was not immediately obvious
in the past.

As demonstrated by our own solar system planets, it is possible that EGPs do
not have the same metal abundances (or relative proportions) as their parent
star.  The main discrepancy between the no-redistribution and $\alpha = 0.5$,
solar abundance, models is the 4.5 \micron\ flux which coincides with the
strong fundamental CO absorption band.  The flux density ratio could be low at
these wavelengths due to enhanced metal abundances.  In order to achieve a CO
absorption feature with flux density ratios as low as in the $\alpha = 0.25$
solar abundance model (which reproduces the 4.5 \micron\ IRAC observations),
the metal abundance would have to be 10 times that of the parent star (see Fig.
\ref{tr2}).  However, given the width of the IRAC band passes, the fluxes
outside the CO absorption band remain high enough to keep the model's
integrated flux density ratio above the 2$\sigma$ error bar.  Note that
increasing the C to O ratio does not improve the comparison to observations.  A
larger C to O ratio increases CO absorption, but simultaneously lowers the
water concentration thereby increasing the planet's fluxes at wavelengths
red-ward of the 4.5 \micron\ that are also included in the IRAC band.

\subsection{HD209458b}

Figure \ref{hd1} shows the T-P profiles for HD209458b assuming solar
abundances,  and the parameters listed in Table \ref{tab1}.  The adopted inner
adiabat was based on evolution calculations which suggest a substantial
intrinsic luminosity for HD209458b's mass and abnormally large radius
\cite[]{Baraffe03}.  The sub-stellar point $\tint = 230$K and the
entropy-matching non-irradiated model for the terminator and night side has
$\teff = 500$K.  

The predicted T-P trend across the day side is similar to that for TrES-1.  The
major differences are due to the greater parent star luminosity for HD209458b,
which leads to a significantly hotter sub-stellar point.  The intrinsic
luminosity is also higher for HD209458b, which leads to warmer regions near the
terminator.  The decline in temperature above the nearly isothermal photosphere
(P $< 0.01$ bar) for the hottest portions of the dayside is consistent with
recent works by other groups \cite[]{Sudarsky03, Fortney05, Iro05}.  This
temperature decline points to some of the differences mentioned above between
the present Rainout models and those based on the earlier Cond approximation.
For HD209458b, with $\alpha = 0.5$, the Cond assumption leads to a nearly
isothermal profile (T $\sim 1700$K) for most of the atmosphere (see Fig. 2 of
\cite{Chabrier04} ).  While the complete removal of TiO and VO via the adopted
Rainout process contributes to the cooler outer atmosphere and hotter
photosphere shown in Fig. \ref{hd1}.  A more detailed comparison between
Rainout and Cond models is given in Barman et al. (2005, in preparation).
 
If clouds are sustainable on the planet's day side, then Fe, \enstatite\ and
\forsterite\ clouds might form at very low pressures across most of the day
side surrounding the sub-stellar point.  Near the terminator, the temperatures
are much cooler than the average temperature across the day side.
Consequently, most of the limb is well below the condensation temperature of
Na, consistent with the findings by \cite{Iro05}.  Condensation of Na near the
limb may contribute to the lower than expected Na absorption detected with HST
\cite[]{Charbonneau02}.

The 24 \micron\ MIPS observations probe the Rayleigh-Jeans tail of the planet's
spectrum, in a region dominated by H$_2$O line opacity.  Figure \ref{hd2} shows
the day side planet-star flux density ratios for the no-redistribution model
and 1-D models with $\alpha = 0.25$ and $\alpha = 0.50$.  At the 1$\sigma$
level, the MIPS observations favor a strong redistribution of the absorbed
stellar flux ($\alpha = 0.25$).  The two hottest cases, no redistribution and
$\alpha = 0.5$, are only marginally in agreement with the 2$\sigma$
observational error bars.  The MIPS observation probes the atmosphere at a
pressure ($\rm P \sim 10^{-2}$) similar to those probed by the IRAC observations.
Note also that for $\lambda > 10$ \micron, the planet's day side spectrum is
nearly identical to that of a non-irradiated brown dwarf with $\teff = 1450$K
(which matches the emergent flux for the $\alpha = 0.25$ case).  The brown
dwarf-like appearance of the spectrum at far-IR wavelengths is to be expected
since, at these temperatures, 24 \micron\ is well within the Rayleigh-Jeans
tail.

Another potentially useful limit for HD209458b has been set at shorter
wavelengths by ground based observations \cite[]{Richardson03, Seager05}. These
observations indicate the planet's spectrum may have a less prominent
2.2 \micron\ peak (or an overall lower luminosity) than predicted by many
models.  This limit also favors $\alpha < 0.50$.

\begin{figure}
\plotone{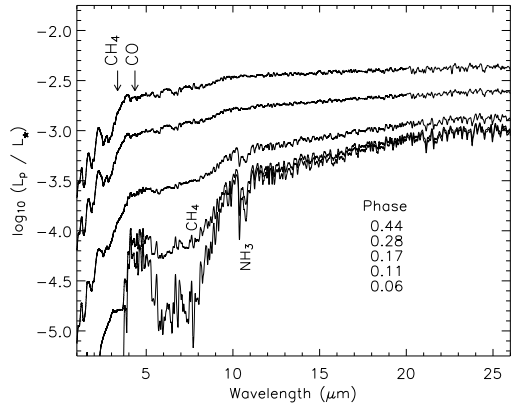}
\caption{Planet-star flux density ratios for HD209458b assuming no
redistribution at five different orbital phases from 0.44 (top curve) to 0.06
(bottom curve) and orbital inclination of 90$^\circ$.  Phase = 0 coincides with
the time of transit and only the night side is visible.  At phase = 0.5 the
planet is behind the star.
\label{ph1}}
\end{figure}

\subsection{Phase-dependent Flux densities}

Since the entire day hemisphere has been modeled by a collection of T-P
profiles and intensity spectra, estimates for the phase-dependent fluxes can be
constructed simply by changing the star-planet-observer orientation and
re-integrating the surface intensities (see Eqs. \ref{fint} and \ref{fsum}).
Figure \ref{ph1} shows the predicted phase-dependent flux density ratios for
HD209458b with no energy redistribution.  Note the significant drop in flux
between 5 and 10 \micron\ and the shift in the peak flux between 1 and 8
\micron\ toward redder wavelengths.  As the phase approaches zero, the coolest
parts of the planet, which are dominated by CH$_4$ absorption, come into view.  

\begin{figure*}
\plotone{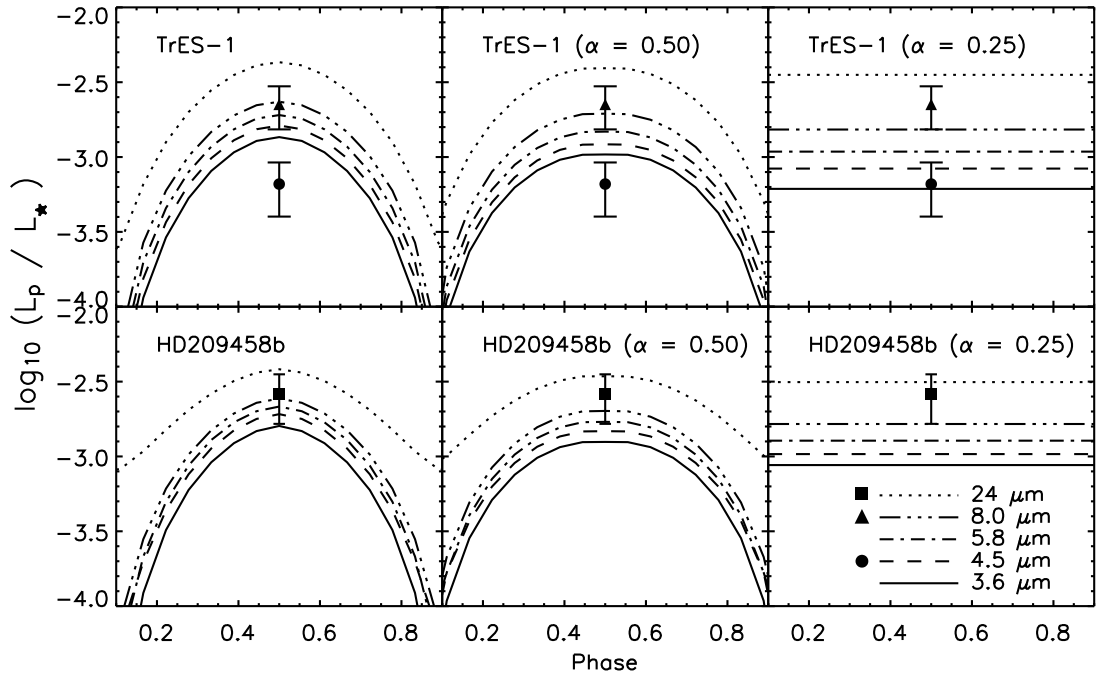}
\caption{Phase-dependent flux density ratios for the 24 \micron\ MIPS band and
the four IRAC bands for TrES-1 and HD209458b.  Model calculations with no
redistribution are shown in the left column.  Results for models with
redistribution characterized by $\alpha = 0.5$ and 0.25 are shown in the middle
and right columns.  The Spitzer observations are shown as filled symbols with
2$\sigma$ error bars. Each data point should be compared to only one of the
theoretical light-curves (see legend for the symbol-linestyle relation).  At
phase = 0.5, only the day side is visible.
\label{ph2}}
\end{figure*}

Flux densities for IRAC and 24 \micron\ MIPS bands are shown in Fig.  \ref{ph2}
for all phases with ($\alpha = 0.5$ and 0.25) and without energy
redistribution.   As the level of redistribution increases, the thermal surface
brightness becomes more uniform and the IR light-curves flatten out. For
$\alpha = 0.25$, the IR light-curves are constant with values equal to those
shown if Fig. \ref{tr2}.  Note that optical and near-IR light-curves will not
be flat when $\alpha=0.25$ due to reflected star light.

A phase shift of TrES-1's hot side would not improve the simultaneous fitting
of no-redistribution and $\alpha = 0.5$ models to the 1$\sigma$ observed 8 and
4.5 \micron\ fluxes for TrES-1.  For HD209458b, both $\alpha = 0.25$ and 0.5
models agree with the 2$\sigma$ error bars but favor $\alpha = 0.25$ at the
1$\sigma$ level.  However, the large 2$\sigma$ error bars do leave room for a
shift of up to 0.25 in phase -- similar to recent results from atmospheric
circulation models \cite[]{Cooper05}.  For both planets, the best fitting model
appears to be one with $\alpha=0.25$.  This may indicate that fast photospheric
winds are present with speeds in excess of 1 km s$^{-1}$ as predicted for
atmospheres with similar day-night temperature differences \cite[]{Cooper05}.
However, using approximations for the radiative and advective time-scales
\cite[]{Showman02, Seager05} and assuming 1 km s$^{-1}$ wind speeds,
$\tau_{rad}/\tau_{adv} < 1$ for $\rm P < 1$bar at the sub-stellar point for
both TrES-1 and HD209458b.  Small values for $\tau_{rad}$ suggest that winds
will not be capable of entirely removing the large day-night temperature
differences even at photospheric depths.

\section{Discussion and Conclusions}

Detailed radiative-convective equilibrium models have been presented above for
the atmospheres of HD209458b and TrES-1.  In the absence of energy
redistribution, the models predict steep horizontal temperature gradients from
the sub-stellar point to the night side that vary substantially with depth.
Compared to earlier works, which assumed very efficient energy redistribution,
the no-redistribution models predict much hotter temperatures across most of
the day hemisphere and significantly cooler temperatures near the terminator.

The existence of steep horizontal temperature gradients in equilibrium models
strengthens the case for strong zonal winds.  However, fast winds ($\sim$ 2 km
sec$^{-1}$) capable of altering the global atmospheric temperature profile are
unlikely to entirely remove these temperature gradients, especially at
pressures lower than a bar \cite[]{Cooper05,Iro05}.  Since $\tau_{rad}$
decreases rapidly with decreasing pressure (i.e. towards the top of the
atmosphere), the horizontal and vertical temperature gradients may well have a
mixture of the T-P characteristics shown above for the various redistribution
scenarios.  For example, the temperature structures could be close to the
static, no-redistribution, scenario at $\rm P < 0.01$ bar where $\tau_{rad} <
10^4$ sec.  Deeper into the atmosphere (where $\tau_{rad} >> \tau_{adv}$)
advection is probably an important mechanism for energy transport which could
lead to a T-P structure similar to that of an $\alpha = 0.5$ or $\alpha = 0.25$
redistribution model for $\rm P > 1$ bar.  Consequently, the T-P profile could,
in some cases, be fairly flat across much of the day side photosphere.  Note
that a flat photospheric T-P profile would produce a spectrum close to that of
a blackbody -- a possibility which, so far, is not excluded by Spitzer
observations (see \cite{Tres1Spitzer} for a blackbody comparison to the TrES-1
Spitzer data).

The 24 \micron\ Spitzer observations of HD209458b have been shown to favor an
atmosphere undergoing efficient day-to-night energy redistribution.  The 4.5
and 8.0 \micron\ observations of TrES-1 also favor an atmosphere which is
experiencing significant redistribution.  However, an $\alpha = 0.25$
redistribution model only agrees with both data points, simultaneously, at the
2$\sigma$ level.  For both planets, it appears that the fully static,
no-redistribution, case is ruled out by the Spitzer data.  However, the data
points are too few and have error bars that are too large to significantly
constrain the model parameters.  

In anticipation of additional Spitzer observations, phase-dependent spectra
have been calculated, along with phase-dependent planet-star flux density
ratios.  For HD209458b, placing limits on the fluxes at the quadrature and near
night side phases (esp. at 24 \micron) would test the large night side
luminosity (with $\teff \sim 400 - 500$K) predicted by evolution calculations.
The ratios of the planet's flux densities at phases 0.5 and 0.25 could also
further constrain the degree to which energy is redistributed to the night
side.

The current work is also applicable to planets in non-transiting orbits.  There
is a direct correspondence between the phase-dependent fluxes shown above (for
an edge-on orbit) and the {\em inclination}-dependent fluxes for a planet at
superior conjunction in an arbitrarily inclined orbit.  Therefore, the
predictions made above suggest that Spitzer could detect flux variations due to
a close-in planet with orbital inclination as small as 45$^\circ$ (i.e., phase
0.375 or 0.625 in Fig. \ref{ph2}).  Since only two of the known transiting
planets orbit stars bright enough to measure the planet-star flux density
ratios, performing similar observations as \cite{HDSpitzer} and
\cite{Tres1Spitzer} for non-transiting planets orbiting nearby stars would be
very helpful.

\acknowledgements
We thank Isabelle Baraffe, Gilles Chabrier and Brad Hansen for their useful
comments and suggestions.  We also thank Dave Charbonneau and Sara Seager for
providing pre-prints of their recent papers and the anonymous referee for
his/her efforts.  This research was supported by NASA through LTSA grant
NAG5-3435 to Wichita State University and Origins of Solar Systems grant
NNG04GL86G to University of California at Los Angeles. We also acknowledge
support by the CNRS. TSB acknowledges additional support by NASA through the
AAS small research grant program and PHH was supported in part by the P\^ole
Scientifique de Mod\'elisation Num\'erique at ENS-Lyon.  Some of the
calculations were performed on the IBM pSeries of the HLRN and CINES, the IBM
SP of the NERSC, the WSU HIPECC, and on NASA's Project Columbia computer
system.  We thank all these institutions for a generous allocation of computer
time.

\end{document}